\documentclass{ifacconf}

\usepackage{graphicx}      
\usepackage{natbib}        
\begin{document}
\begin{frontmatter}

\title{Characterizing Quality of Experience for Demand Management in South Brazil \thanksref{footnoteinfo}} 

\thanks[footnoteinfo]{The authors thank CNPq for projects 401126 /2014-5, 303702/2011-7 and 305785/2015-0.}


\author[First]{Jos\'{e} Diogo Forte de Oliveira Luna} 
\author[First]{Guilherme Nascimento Gouv\^{e}a dos Reis} 
\author[Second]{Paulo Renato da Costa Mendes}
\author[First]{Julio Elias Normey-Rico}

\address[First]{Renewable Energy Research Group, Dept. of Automation and Systems Engineering, Federal University of Santa Catarina, Florian\'{o}polis, Brazil. (e-mail: jose.luna@ifro.edu.br, guilherme.gouvea@grad.ufsc.br, paulo.mendes@itwm.fraunhofer.de, julio.normey@ufsc.br).}
\address[Second]{Fraunhofer Institute  for Industrial Mathematics, Kaiserslautern, Germany.}

\begin{abstract}                
The present work delivers the results of a survey conducted in Florian\'{o}polis, on the southern region of Brazil, using a digital questionnaire which inquired about the interaction with demand management systems (DMSs) in a smart house context. In particular, the survey addressed the interviewees' thoughts on demand management and gathered data to enable the design of Quality of Experience (QoE) based demand management policies. The investigation of QoE-aware approaches is significant as it enables the DMS to take decisions not only based on economic metrics, but also taking the discomfort caused to the users into account. The number of responses guaranteed a confidence level of 95\% and an error margin of 4.63\% and the content of such responses showed favorable disposition of the interviewees on allowing interventions of DMSs on their energy consumption habits as long as it reduces their expenses with energy. The data concerning discomfort caused by demand management actions on several home appliances was treated using clustering techniques and different typical user profiles were leveraged, represented by the centroid of each cluster. Those profiles can be used to deploy QoE-aware DMSs tailored for the local reality, which seem as a promising business model as smart houses become a reality.

\end{abstract}


\end{frontmatter}

%
\section{Introduction}





The stable operation of any grid requires the balance between the power generation and its consumption. Maintaining this balance, specially in peak hours, require a series of mechanisms that usually include incentive to consumers to participate in demand side response \citep{eissa2019, Kirby2003}. 

As the penetration of information and communication technology increase associated with distributed renewable sources, energy storage systems (ESS) and smart loads, modern buildings are becoming complex cyber-physical systems that can be understood as microgrids (MGs). In this scenario, energy management systems (EMSs) that integrate and coordinate the distributed generation (DG), the ESS and the loads are able to contribute to well being of the grid by optimizing the energy consumption of the household by scheduling the usage of energy storages and by implementing demand management (DM)  \citep{luo2019}.

DM techniques aim to flatten the daily demand curve, increasing the demand on low consumption hours and reducing the demand on more heavily loaded hours, without compromising the operation of the appliances in the building.
This implies in energy cost reduction for the household owners once the energy tariff is more expansive during high demand periods and cheaper on light load hours \citep{Kirby2003}. 
DM actions are realized by means of simply curtailing part of the consumption of a given load \citep{Li2016}, shedding it completely \citep{garifi2018} or shifting a load on the time schedule \citep{zhou2012qoe}, meaning that it is possible to run such load earlier or later than originally intended by the user. As not every load is eligible to DM action \citep{Shoreh2016}, practical realization of DM policies usually classify loads on their capacity of being shifted, interrupted or curtailed.


Nevertheless, even when accounting for shiftable, interruptible and curtailable loads only, DM policies may cause discomfort on the users \citep{zhou2012qoe}. While impacts on energy consumption can be easily measured in financial terms, the discomfort due to interruption or change in the pattern of such consumption can not be as easily measured. As stated in \cite{floris2018}, while many solutions have been proposed in the last years due to the widespread diffusion of local communication technologies and embedded systems, their developments are mostly focused on reducing costs, achieving system reliability and other technical criteria while only limited attention has been put to the impact of the system performance on the user. This is somehow counter-intuitive as humans are the final recipient of the service, but can be deemed to the lack of ways to quantify that human experience.
While it is common to consider penalties for load curtailments when DM acts, even recent works, such as \cite{garifi2018,Zhang2018}, do not present a clear way in which those penalties represent the inconvenience caused to the users. 

\subsection{Literature Review}
 
In order to avoid arbitrary computation of penalties for load shift and curtailment, the adoption of a Quality of Experience (QoE) metric for demand scheduling in a MG environment was proposed in
\cite{zhou2012qoe}. The QoE is a metric that can be used to gauge the impact of DM actions on the satisfaction of the user. QoE is defined as the degree of satisfaction and comfort that a user experiences with a certain application or service, resulting from the ratio of accomplishment of their expectations for that particular utility. It is an evaluation of human experience when interacting with a technological entity in a certain context \citep{zhou2012qoe}. According to \cite{Pilloni2018}, QoE is commonly a subjective evaluation performed on a group of people in order to evaluate a certain application or service. The measurement of QoE is, then, used to enhance such application or service. On a demand management system context, interaction between the users and the system is assumed. Such interaction will allow the user to select preferences and train the system to comply to those, enabling the system to find a suitable trade-off between reducing the energy cost and increasing the QoE.

In a MG context, QoE is related to three factors: utility, accessibility and customization. Utility is the capability of satisfying the needs and desires of the user and how much they perceive the service as valuable. Accessibility describes the practice of allowing all users to handle the services and applications equally, despite their skill levels or deficiencies.  Customization is the capability of providing contents and services that are tuned to each individual based on previous knowledge of their preferences and behavior \citep{zhou2012qoe}.

QoE has been largely used in telecommunications and computation contexts, but never before for energy management. Despite being the seminal work of QoE-Aware Demand Management, \cite{zhou2012qoe} did not effectively establish a QoE metric for DM, as the study focuses on defining the network architecture and the overall communication framework necessary for the implementation of a QoE-aware smart home energy management system, using an arbitrary logarithmic function as an example on the text.

To the authors best knowledge, the theme remained untouched until 2015, when \cite{Floris2015} proposes an effective metric for QoE based on specific equipment. This metric describes the discomfort caused by the shifting in time of each one of those equipment. A subjective survey was conducted and each interviewee was asked about their usage of the following home appliances: washing machine, dishwasher, clothes dryer, electric oven, micro-wave oven, air conditioning and water heating. In particular the subjects were asked about the annoyance due to anticipating and postponing the usage of those equipment in half, one, one and a half, two, two and a half and three hours. The discomfort caused by such shift was quantified in a scale from 1 to 5, with 1 being the least annoying and 5 being the most annoying. After collecting the answers, a k-means approach was used to come up with three to six different usage profiles for each home appliance. For each profile of each equipment, then, was obtained a QoE curve for the anticipating and postponing of usage. Based on that, an algorithm for smart house energy management, based on if-then rules, was proposed in order to reduce the energy cost, making better usage of renewable generation, without exceeding determined QoE acceptable limits. The conducted survey interviewed 64 people in Italy. No information is presented on the work about the confidence level or error margin of the survey, which is reasonably desired when seeking to characterize a given universe.

Two years later, in 2017 \cite{Li2017} categorizes the loads in three groups: delay tolerant, flexible and delay intolerant and inflexible and delay intolerant. Delay tolerant loads may be shifted in  time to shave demand peaks. Flexible and delay intolerant loads are those that cannot be shifted in time, but can have its power controlled, such as Heating, Ventilation and Air Conditioning (HVAC) applications or electric vehicles (EV) charging. Building upon these considerations, the work uses the curves obtained in  \cite{Floris2015} and finds a piece-wise exponential regression to describe the relationship between the delay and the QoE. After that, an algorithm for smart house energy management, also based on if-then rules, is presented, taking into account the variable energy price and the QoE.
The work presents simulation results indicating better performance, caused by the adoption of the proposed technique.

Afterwards, it was proposed in \cite{Pilloni2018} a revisited version of their previous work, including new considerations on loads that are not shiftable but have flexibility, similarly to \cite{Li2017}. The work considers the available renewable generation in order to allocate the loads, while respecting QoE thresholds, with a new algorithm which, also, is based on if-then rules.

In the same period, \cite{Li2018}, revisit their previous approach, enhancing their work. On this new study a variable QoE threshold is proposed, in contrast with the fixed threshold that had been used so far. In this approach a fuzzy logic controller was employed for dynamically setting the QoE threshold, designed to resolve the trade-off problem between the QoE degradation and reduction of peak load and electricity bill. The controller receives the power consumption and the deviation from the average power as inputs in order to calculate the limit to what the QoE will be allowed to fall before the appliance being started despite the energy cost. Once again the algorithm is based on a if-then rule set that is triggered when a request for appliance start arrives. The appliance can either be started right away, be scheduled to start after a certain delay or, in case of HVAC, be started with a fraction of the total power.

Later in 2018, \cite{floris2018}, build upon their last work, where the control system take decisions considering both the user profile and the cost. This time there are two algorithms acting together: a cost saving appliance scheduling algorithm that is aimed at scheduling controlled loads based on users' preferences and electricity price and a QoE-aware renewable source power allocation algorithm that modifies the working schedule of appliances whenever a surplus of energy has been made available by the renewable generation. The authors focus on the identification of the user profile, initially adopting one of the profiles obtained from the survey and building it from there, based on the feedback from the user. This work included a practical experiment that involved 12 subjects considering their interaction with the energy management system comprehending the usage of washing machine, dishwasher, clothes dryer and electric oven. Using the proposed learning strategy, the best fitting profile was found after two feedback sessions, in average.


The potential of QoE as a user comfort measures holds specially when contrasted with other approaches. In 2016 \cite{christensen2016} performed a comparison between Analytic Hierarchy Process (AHP), Discrete Choice Modeling (DCM) and Simple Multi-Attribute Rating Technique Exploiting Ranks (SMARTER). AHP is based on pairwise comparisons using a predetermined scale, user responses are used to calculate local and global priorities and rank the alternatives. In DCM the users rank a finite set of alternatives from most to least desirable, within a series of choice situations. 
Both approaches, in the end, resort to asking users a number of questions to quantify which situations are more desirable, ranking them and inferring weights of a objective function from that. SMARTER is more direct, asking users to rank attributes and weights are inferred from that. In \cite{edwards1994}, the authors of SMARTER, claimed that ``In short, when weights don't pick the best option, the one they do pick isn't too bad''. QoE, on the other hand, allows a different approach as comfort degradation is described for each appliance, so not only is possible to assess if some appliance is preferred, but also assess the overall evolution of the comfort degradation over time.

\subsection{Contributions}

That said, while there are currently few works concerning the usage of QoE on DM, the decrease in the interval between publications on the subject reflects a promising technique that can find meaningful usage in a MG context. In particular, considering that the existing literature used the results of the QoE survey performed by \cite{Floris2015} in Italy, the lack of other results from different locations arise as a evident gap that must be addressed, both for the sake of comparison and localization: adapting the proposed techniques to specific target markets that have different cultures and, presumably, perceptions of an interaction with a EMS. 

In spite of that, the present work aims to perform a survey in Florian\'{o}polis, on the southern region of Brazil, examining its results concerning the acceptability of a QoE-aware demand management system on smart houses with a confidence level of 95\% and an error margin of 5\%. This survey adapts the questionnaire applied by \cite{Floris2015} to the local reality, changing the list of home appliances to the ones that are more adequate for a smart home scenario in Brazilian territory. The results of this survey enable the implementation of QoE-aware DM policies on a local basis and also enable a comparison between the results of similar surveys in different areas of the world. Moreover,  the survey shows opening for new business models, incorporating QoE-aware demand management to energy management systems, aggregating value to smart houses and thus creating a gap for new business models on the local market.
Finally, complementary data relating to the education level, age, family income bracket are also analyzed to provide insight on how different groups of society perceive the possibility of interaction with a QoE-aware DM system and smart houses in general.


In order to do so, the text is divided as follows: the second section contain a brief explanation on the energy sector, aiming to provide context to the reader. The third section presents the methodology, with special attention to the survey contents and the tools used to analyze the responses. The fourth section examines and discusses the results of the survey. Finally, the fifth section finishes the work with some conclusions and perspectives drawn from the obtained results.

\section{Brazilian Energy Sector Overview}

The Brazilian Energy Sector is regulated by the National Electric Energy Agency (ANEEL, from the Portuguese Ag\^{e}ncia Nacional de Energia El\'{e}trica). Accordingly to its regulations, the consumers are broadly classified in two groups: A and B. Group A covers the consumers supplied on voltages equal or over 2.3 kV which commonly are industries and large commercial sites. Group B, then, encompasses consumers supplied on voltages lower than 2.3 kV and it is further divided into four subgroups. Subgroup B1 for residential consumers, B2 for rural consumers, B3 for small commercial or industrial buildings and B4 for public artificial lighting \citep{aneel414}. 

These subgroups that form group B branch down to subclasses. Subgroup B1 is divided into normal residential and low income residential, that is connected with the governmental program Social Tariff, that benefits low income families \citep{lei12212,decreto7583}. Subgroups B2 is divided into non-cooperative rural, electrification cooperative  and public irrigation service \citep{aneel449}. Subgroup B3 is divided into commercial, non-electrical transport services, communication and telecommunication services, philanthropic associations and entities, religious temples,  condominium administration, highway lighting, traffic monitoring, and other services \citep{aneel414}.

\begin{figure}[htbp]
\centerline{\includegraphics[scale=.95]{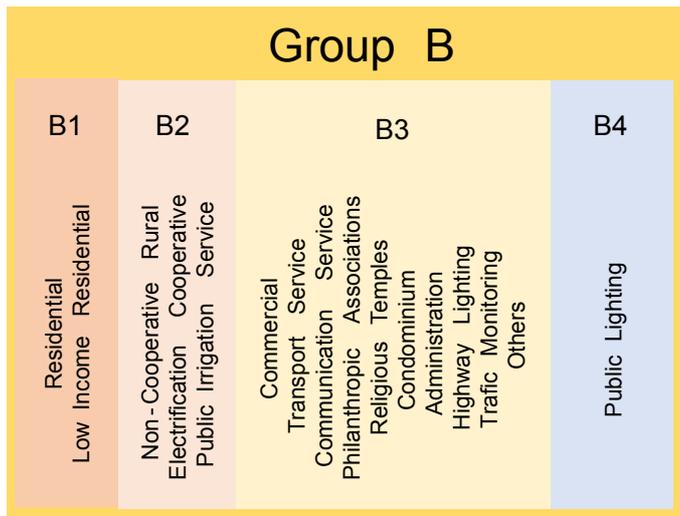}}
\caption{Diviion of Group B consumers.}
\label{QoE32}
\end{figure}

\begin{figure}[htbp]
\centerline{\includegraphics[scale=.6]{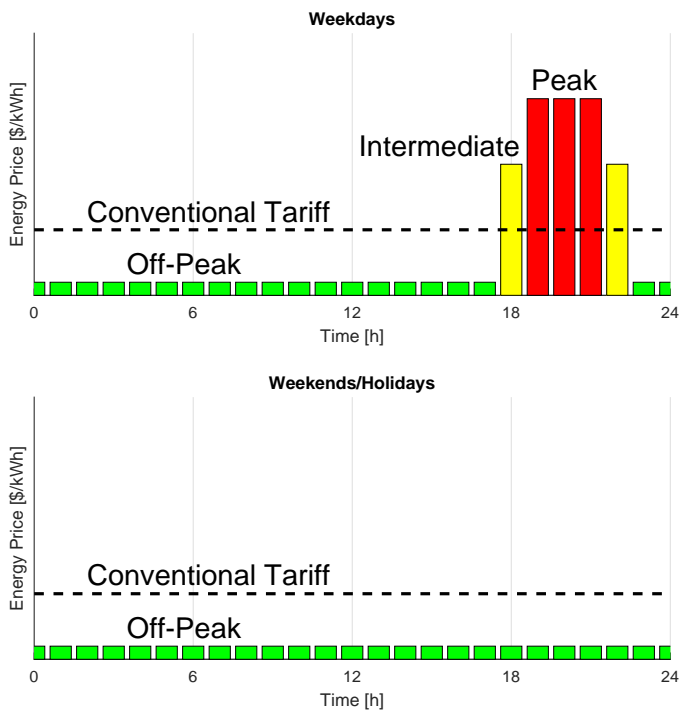}}
\caption{Relative prices for the conventional tariff and the white tariff. Red: peak hours, yellow: intermediate hours, green: off-peak hours, black: conventional tariff. Adapted from: \cite{tarifabranca}.}
\label{QoE31}
\end{figure}

Group B consumers have been charged with the conventional monomial tariff. This tariff charges active energy consumption with a flat price, independently on the hour of the day \citep{aneel479}. However, since 2018,  group B consumers, except for subgroup B4 and low income subclasses of subgroup B1, may opt to be charged by the white tariff. This tariff is characterized by three different values for energy consumption accordingly to the time of the day. These times are designated as peak hours, where the largest demand is concentrated, intermediate hours, one hour before and one hour after the peak hours, and off-peak hours, with the lesser energy demand. In particular, the peak hours are three consecutive hours that each local energy distribution company must elect reflecting their demand curve. Weekends and holidays are considered fully off-peak hours \citep{aneel733}. Figure \ref{QoE31} illustrates the the price profile of the two options.

The tariff the consumer is charged is composed by two basic terms the Tariff for Usage of Distribution Systems (TUSD, from Tarifa de Uso do Sistema de Distribui\c{c}\~{a}o) and the Energy Tariff (TE, Tarifa de Energia). The TUSD, itself, is composed of several components:
\begin{itemize}
    \item TUSD - B wire, which corresponds to the cost of service provided by the distribution company itself;
    \item TUSD - A wire, which represents the cost of usage of third party distribution or transmission networks;
    \item TUSD - charges, which is the cost of taxes and charges related to the energy distribution service;
    \item TUSD - Technical Losses, which encompasses technical losses;
    \item TUSD - Non-Technical Losses, which encompasses non-technical losses;
    \item Fuel Consumption Bill (CCC, Conta de Consumo de Combust\'{i}veis), that addresses the cost of thermoelectric generation for isolated systems and low hidraulic generation periods;
    \item Energetic Development Bill (CDE, Conta de Desenvolvimento Energ\'{e}tico), a tax aimed for the energetic development of the federative units;
    \item Incentive Program for Alternate Electric Energy Sources (PROINFA, Programa de Incentivo a Fontes de Energia Alternativas), which aims to improve the participation of renewable sources on the national energy matrix.
    \end{itemize}  
The TE, on the other hand, is composed of two parcels: TE - energy, that covers the cost of energy consumption itself and TE - charges,  which is the cost of taxes and charges related to the energy consumption.


\subsection{Distributed Generation}

Distributed generation is allowed to access to the distribution networks since 2012, if they are classified as microgeneration or minigeneration and make usage of renewable sources \citep{aneel482}. Microgeneration is classified as any renewable generation with installed power equal or less than 75 kW, while minigeneration has installed power higher than 75 kW and lesser or equal to 3 MW, for hydraulic sources, and 5 MW for other renewable sources \citep{aneel687} . For group B consumers, the installed generation power is limited by their installed load \footnote{Which is agreed with the local distribution company when the consumer signs the adhesion contract}, as the primarily function of distributed microgeneration is understood as self-consumption \citep{aneel482}. 

When the generation exceeds the consumption, the excess power is supplied to the grid and accounted by the bidirectional metering. While the consumer is unable to sell this energy it is accounted as energy credits, that the user can consume later, in a maximum period of 60 months. The generated credits are equal to the value that would be payed by the consumer if they would consume the same amount of energy, in a way that the network acts as a virtual storage for the consumer. Even so, there is a minimum availability fee that the consumer has to pay, even if the energy credits generated cover all of his or her consumption \citep{aneel482}.

Additionally, there are three modalities of distribution generation introduced in 2015  \citep{aneel687} that are relevant for the discussion in a MG context: enterprise with multiple consuming units, remote selfconsumption and shared generation. An enterprise with multiple consuming units is characterized by the independent usage of energy by different consumers on the same building or in contiguous buildings whose total distributed generation is split between the consumers. Remote selfconsumption describes the possibility of a user who owns more than one consuming unit supplied by the same distribution company to use the energy credits from the excess generation of one of their consuming units to compensate the consumption of another consuming units. Finally, shared generation is a combination of a enterprise with multiple consuming units and remote selfconsumption, in the sense that a group of consuming units with different owners, associated in a cooperative or consortium, are able to compensate their consumption with credits originated from the generation of a non contiguous site.

The regulations, however, are currently being revised. ANEEL opened the public audience 01/2019 on the evaluation of regulatory impact on the changes on the current regulation. The main proposed change is on the compensation of the excess generation. Currently the all of the tariff is compensated. It is argued that if the user that have their own distributed generation do not pay for the usage of the distribution network, it will be solely payed by the remaining users, while it is used by all of the consumers. On the other hand, generated distribution has significative benefits for the enviroment and the operation of the overall network and the current regulation favor the growth of the distributed generation market. That said, the current proposal is, for local compensation, that, when the total Brazilian distributed generation reaches 3.5 GW\footnote{ANEEL predicts it will happen in 2025 \citep{consulta012019}.}, the compensation stops covering the TUSD - wire B, meaning that the consumer would still pay for the transport cost of the consumed energy. That would mean that only 72\% of the excess generation would be compensated \citep{consulta012019}.

\subsection{Energy Management}

On a white tariff scenario, in order to reduce the consumer individual electricity bill, the most part of the load should be averted from the peak and intermediate periods, shifting the operation time of appliances that would run on these hours to other periods as much as possible. Additionally, considering the existence of distributed generators, in particular, photovoltaic ones, on the smart house, these loads may be shifted to hours where the generation surpass the consumption as the change in the regulations proposed on \citep{consulta012019} make it more profitable to consume directly the excess generation than to virtually store it as energy credits in the compensation system. 

In that context, the simple usage of a grid tie inverter is not sufficient to provide optimal usage of the renewable generation. An EMS is required to perform DM actions in order to fit the load curve as better as possible to the generation curve. In that regard, QoE-aware demand management poses itself as a promising alternative, as it allows the EMS to act guided by a compromise between purely minimizing financial costs and caring for the satisfaction of the residents of the smart house.

\section{Methodology}


Aiming to portray a local profile on the perception and behavior with respect to DM policies implemented  on a EMS, a questionnaire was developed encompassing questions on the acceptability of demand management itself and its application on certain equipment. While the survey is inspired on the one applied in \cite{Floris2015}, the present questionnaire was tailored to local characteristics. In particular, the survey targeted students and faculty of the Federal University of Santa Catarina. Considering the local population of 492,977 inhabitants\footnote{2018 estimate of the Brazilian Institute of Statistics and Geography (IBGE).} a sample of 384 people assures an error margin of 5\% and a confidence level of 95\%. The size of the sample, $n$, is calculated as:

\begin{equation}\label{eq1}
    n = \frac{NZ^2p(1-p)}{(N-1)e^2+Z^2p(1-p)},
\end{equation}
where $N$ is the size of the total population, $Z$ is 1.96 for a 95\% confidence level, $e$ is the error margin and $p$ is the estimated proportion of the population, adopted as 0.5 for the worst case scenario.

The questionnaire had two main sections, one for identification of the participant and another about possible interactions with a demand management system. 
In the identification section data like educational level, age group, sex and family income bracket were leveraged as to enable a social, economical and cultural profiling of the interviewees.
In the technical section, the interviewee were asked if they would be comfortable in letting a demand management system interfere in the start time or power consumption of certain home appliances, such as washing machine, clothes dryer, dish washer, electric shower, HVAC and electric vehicle charging, if that meant a reduction on their energy bill. If they agreed on having a demand management system intervening in their energy consumption habits, they were prompted to rate the eventual annoyance of having those appliances start time anticipated or postponed in a certain time, or, if the case, having their consumption reduced.

It was asked if the interviewee would be willing to participate in future questionnaires about the same subject, as there was intentions to understand better the consumption profile of the region's population.


In accordance with the local regulation, the proposed survey is sustained by the article 1, first paragraph, subsection I, of the Resolution 510/2016 of the Plenary of the National Health Council, which excludes public opinion surveys with unidentified subjects from the requirement of approval by the Research Ethics Committee.

\subsection{Survey}

The questionnaire was applied digitally, using the Google Forms platform and was distributed on e-mail lists of the campus. The survey started with a introductory text, explaining briefly the concepts of smart houses as well as energy and demand management \footnote{The original survey, in Portuguese, is available in: https://goo.gl/forms/izlia0WQZz825f1A2}. 

The first section aimed to stratify the interviewee. The first question asked the higher completed educational degree the interviewee had: fundamental, high school, technical school, graduate or post-graduate \footnote{Brazilian Educational System is organized in four levels: fundamental and high schools, technical and higher educations. The first is for children between the ages 6-14, the second for teens between ages 14-17. Technical education usually substitutes high school and is meant to prepare students to work as technicians. Higher education is optional and grants formal knowledge.}. After that the interviewee assigned to each age group they belong to: under 18, between 18 and 25, between 26 and 35, between 36 and 45, between 46 and 56, between 56 and 65, or over 65. Following, the interviewee was prompted to state their sex and their family income bracket: less than two minimum wages \footnote{The Brazilian monthly minimum wage during the period of the survey was BRL 954.00, about EUR 220.00.}, between two and five minimum wages, between five and 10 minimum wages, between 10 and 20 minimum wages or above 20 minimum wages \footnote{These values are used by IBGE to perform economic stratification in classes E, D, C, B and A.}.

The interviewee were also asked about their relationship with the University: if they were students, faculty members or had other relationship with the institution, and what \textit{campi} did they study or work at. 

On reaching the technical section of the survey the interviewee was presented to a brief explanation of the white tariff to give additional context to the scenario in which the energy and demand management system would be employed. 

The survey asked if the interviewee would feel comfortable having an EMS intervening on their energy consumption, if they would feel comfortable having an EMS shifting loads and if they would feel comfortable having an EMS shedding loads, if those actions would result in reduction on their energy tariff. It is also asked if they would agree to allow load shedding if they would receive financial compensation beyond reduction on the tariff.

This set of questions aim to evaluate the general disposition of the interviewee to having an EMS employing DM actions by shifting or shedding loads. The last question specifically address the DM responses as service. 

Next, the interviewee were asked about the intervention of the EMS on specific home appliances, inquiring about the acceptability of shifting the running time of washer machine, clothes dryer and dishwasher and shedding or curtailing HVAC equipment, electric shower and applying a slower charge on a battery of a electric car. For this section, the users were asked to make an exercise of imagination if they currently do not own some of these appliances. For each of these equipment it was asked whether they would accept the action of the EMS and, if so, to rate how annoying would be depending on the intensity of the action. The rating was from 1, meaning not uncomfortable, to 5, meaning highly uncomfortable. This rating was applied to anticipating or postponing the starting time of washer machine, clothes dryer and dishwasher by half an hour, one hour, two hours, three hours and four hours. The rating was also applied to increasing room temperature by one 
to five degrees, on summer, decreasing room temperature by one 
to five degrees, on winter, decreasing water temperature by one 
to five degrees, on the electric shower, and extending the charging time of the electric car by half an hour, one hour, two hours, three hours and four hours.

\begin{figure}[htbp]
\centerline{\includegraphics[scale=.43]{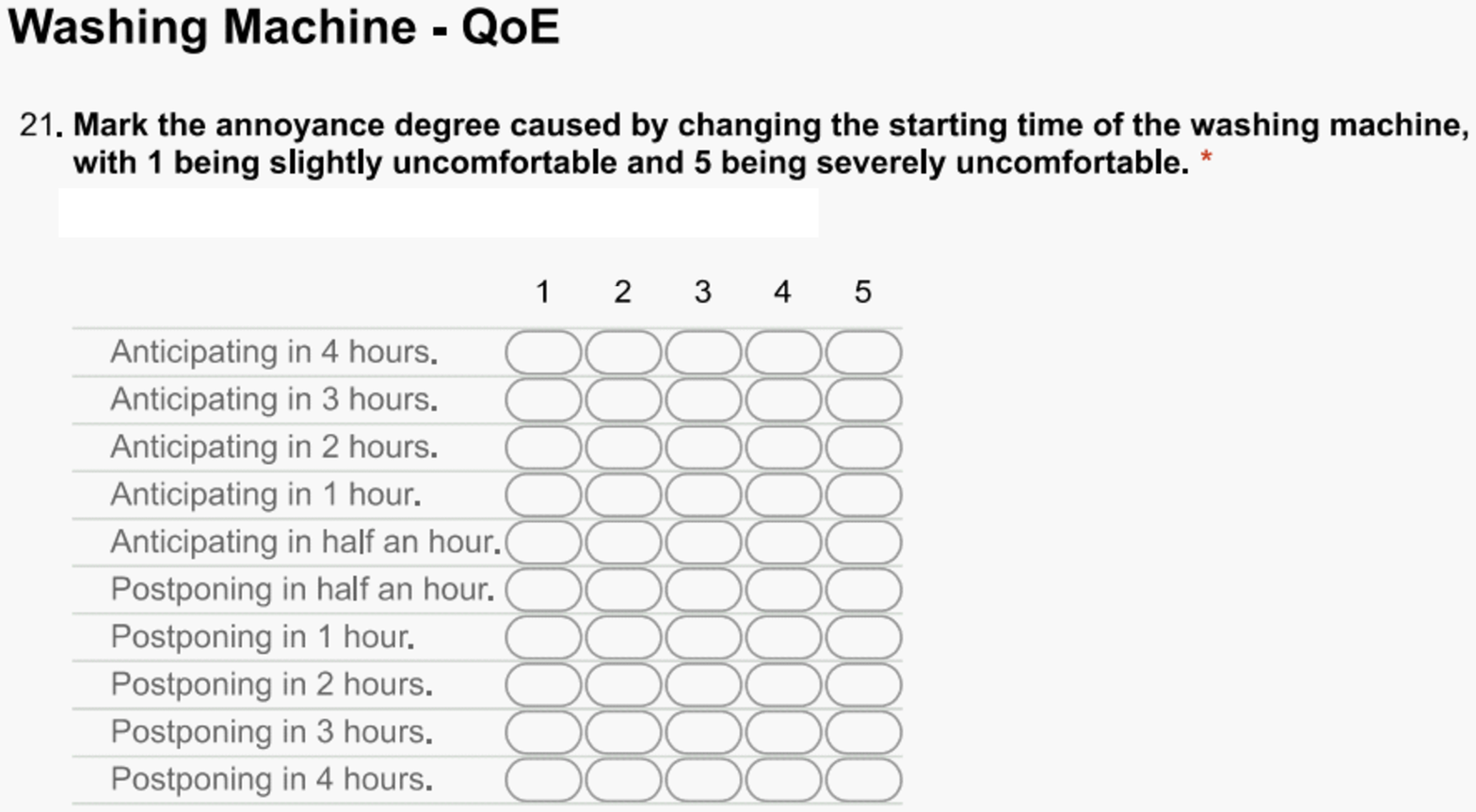}}
\caption{Example of question asking the interviewee to rate the degree of annoyance resulting from the intervention of a EMS.}
\label{QoE34}
\end{figure}

On figure \ref{QoE34} a sample question is shown to illustrate these later questions. The interviewees were presented to a grid with the length of the anticipation or delay on the rows and the intensity of the caused discomfort on the columns and were prompted to check one value for each row. The same scheme was used for questions concerning changes in the temperatures.

Additionally, the interviewee were also inquired if they would agree if the EMS reduced illumination to adequate levels on rooms currently being used and if they would agree to the EMS warning them if the bath using an electric shower was taking longer than a set threshold. Finally, the users were asked if they would like to answer a future questionnaire on electric cars, and if they had any comments or suggestions.


\subsection{K-means Algorithm}

The k-means algorithm is a popular method for cluster analysis, widely known in data science and used to group observations into k clusters \citep{jain2010data}. Each one is composed by the observations close to its mean.

Given an initial set of representatives (k guesses for the means), the algorithm iteratively sort the observations by computing which mean is closer (for some metric) to each observation. Then, the means are recalculated (given the clusters were modified) and these steps are repeated until some condition is reached. It is noteworthy that representatives not necessarily occur in the observations.

In this work, the method is employed to determine representatives of interviewees and these will be selected as characteristic consumers profiles to the EMS.

\subsection{Elbow Method}

As for the k-means algorithm the number of clusters, namely k, is an input, the problem of deciding its value arises. Clearly, choosing k as the number of observations would describe perfectly the data: zero variance in each group and no loss of information. However, this would be of no profit, the data analysis would be computationally expensive. On the other hand, for small k the clusters wouldn't represent the data satisfactorily. So, even with low computational costs, the data analysis, on to the clusters, would be very inefficient.

The trade-off between computational load and fidelity to the original data can be measured by the decrease of total within sum of squares (WSS), i.e., the effect on the squared error by adding one more group. The elbow method tries to select the largest number of clusters that effectively affect on to data modeling. As the name suggests, it looks for the optimal k graphically: where the curve of WSS by number of clusters bends.

\section{Results and Discussion}

Having completed the survey, the obtained results are examined below.

\subsection{Statistical Analysis}

The survey was applied from December of 2018 to April of 2019, having received 448 responses, granting a confidence interval of 95\% and an error margin of 4.63\%, accomplishing the intended values. The resulting error margin was calculated using Equation \ref{eq1}.


From the results of the identification section, 23.39\% of the respondents have completed the secondary school, 8.06\% have completed a technical school, 27.82\% are graduated and 40.82\% already have a post-graduate degree. About the age groups, 0.60\% are under 18, 43.75\% are between 18 and 25, 35.89\% are between 26 and 35, 10.48\% are between 36 and 45, 5.85\% are between 46 and 55, 3.26\% are between 56 and 65, and 0.20\% are above 65 years old. Regarding 41.73\% of the interviewees are females and 58.27\% are males. Concerning the income brackets, 13.06\% have a family income equal or lower than two minimum wages, 32.45\% receive between two and five minimum wages, 28.98\% have a family income between five and ten minimum wages, 17.76\% of the respondents receive between ten and 20 minimum wages and 7.76\% have a family income greater than 20 minimum wages. 

With respect to the interviewees role, 81.41\% are students of the university, 3.64\% are part of the administrative staff, 5.45\% are part of the teaching staff and 9.49\% are part of the local community, not directly related to the university.

As the survey was conducted on college grounds the high numbers of highly qualified interviewees was expected. Additionally, Florian\'{o}polis is altogether one of the cities with higher concentration of qualified people, for instance, 24.18\% of the population with ages above ten have completed superior educations, while the national average is 8.31\% \footnote{information obtained from IBGE (2019).}, it has 533.44 PhDs for each 100,000 inhabitants, while the total Brazilian number is 104.83 PhDs for each 100,000 inhabitants \citep{doutores}.

The age distribution of the participants was also expected, as the majority of the respondents are graduate and undergraduate students. This scenario also explains the income bracket profile of the respondents.


Concerning the direct questions, 93.95\% of the interviewees answered they would feel comfortable having an EMS intervening on their energy consumption if that implied in reduction on their energy bill. Also, 90.32\% of the interviewees (93.13\% of those who would accept the interference of an EMS)  answered they would feel comfortable having an EMS shifting loads and 71.37\% (74.89\% of those who would accept the interference of an EMS) would feel comfortable having an EMS shedding heavy loads momentarily. Nonetheless, 24.33\% of those who would accept load shifting would not accept load shedding. Additionally, 84.27\% of the interviewees would accept an EMS shedding heavy loads momentarily if they would have financial gains beyond reduction on their energy bill.  From the group that would not accept load shedding just for reduction on their energy bill, 47.18\% would change their mind if they would receive financial compensation. 

Considering the actuation of the EMS on the artificial lighting, 88.11\% of the interviewees would like the EMS to reduce illumination to adequate levels. Moreover, 87.90\% would like to be warned if their bath took longer than a set time threshold.

The results show that the interviewees mostly have an accepting view of the scenario of interacting with a EMS if that lead to more efficient consumption. Furthermore, people seem more comfortable with the idea of the EMS shifting load than shedding it momentarily. 

Almost half of those who said they would be uncomfortable with a EMS shedding loads momentarily would change their mind if they would have any financial compensation for that. The fraction of people who would accept this kind of interaction indicates a favorable scenario for demand response as a service schemes. While Brazilian regulation is still stepping towards demand response for Group A consumers \citep{aneel792,relatorioDR}, the provision of load aggregators may pave the way to virtual power plants where multiple consumers can negotiate demand management actions and be rewarded by that.

The answers for the objective questions and their relation to the segments quantified on the identification section are presented bellow. The questions are identified by numbers, as follows.

\begin{enumerate}
    \item `` Would you feel comfortable having a smart device managing and interveining on your energy consumption if that resulted in reduction on your energy bill?''
    \item `` Would you feel comfortable having a smart device with authorization to anticipate or delay the running time of domestic appliances if that resulted in reduction on your energy bill?''
    \item `` Would you feel comfortable having a smart device with authorization to turn off momentarily high consumption domestic appliances, like heaters or air conditioning, if that resulted in reduction on your energy bill?''
    \item `` Would you feel comfortable having a smart device with authorization to turn off momentarily high consumption domestic appliances, if there was financial compensation for that beyond a reduction on your energy bill?''
    \item `` Would you feel comfortable having a smart device reduce the exceeding artificial lighting  to adequate levels, if that resulted in reduction on your energy bill?''
    \item `` Would you feel comfortable having a smart device warn you if the programmed bath duration was exceeded?''

\end{enumerate}

Although it was not a question, the number 7 is used to indicate the number of people who answered negatively to question 3 but would change their mind if they would receive financial compensation and answered positively on question 4.


On table \ref{tabpestudo} the answers for questions 1 to 7 are segmented by educational level. As seem from the data, the overall acceptance of DM policies is wide spread among the different levels, and do not change significantly.

\begin{table}[htb]
\caption{Answer of the objective questions. S.E.: secondary education, T.E.: technical education, G.: graduate, P.G.: post-graduate.}
\begin{center}
\begin{tabular}{|c|c|c|c|c|c|}
\hline
\multicolumn{2}{|c|}{} & S.E. & T.E. & G. & P.G. \\ \hline
 {1} & Yes & 93,10\% & 90,00\% & 96,38\% & 93,56\% \\ \cline{2-6} 
 & No & 6,90\% & 10,00\% & 3,62\% & 6,44\% \\ \hline
 {2} & Yes & 88,79\% & 87,50\% & 95,65\% & 88,12\% \\ \cline{2-6} 
 & No & 11,21\% & 12,50\% & 4,35\% & 11,88\% \\ \hline
 {3} & Yes & 65,52\% & 77,50\% & 78,26\% & 68,81\% \\ \cline{2-6} 
 & No & 34,48\% & 22,50\% & 21,74\% & 31,19\% \\ \hline
 {4} & Yes & 81,90\% & 95,00\% & 89,13\% & 80,20\% \\ \cline{2-6} 
 & No & 18,10\% & 5,00\% & 10,87\% & 19,80\% \\ \hline
 {5} & Yes & 89,66\% & 85,00\% & 88,41\% & 87,62\% \\ \cline{2-6} 
 & No & 10,34\% & 15,00\% & 11,59\% & 12,38\% \\ \hline
 {6} & Yes & 91,38\% & 92,50\% & 90,58\% & 83,17\% \\ \cline{2-6} 
 & No & 8,62\% & 7,50\% & 9,42\% & 16,83\% \\ \hline
 {7} & Yes & 17,24\% & 17,50\% & 11,59\% & 11,88\% \\ \cline{2-6} 
 & No & 82,76\% & 82,50\% & 88,41\% & 88,12\% \\ \hline
\end{tabular}
\label{tabpestudo}
\end{center}
\end{table}


On table \ref{tabp17eidade} the answers for questions 1 to 7 are segmented by age. The results for groups under 18 and above 65 were omitted as the number of respondents for these groups were too small.

\begin{table}[htb]
\caption{Answer of the objective questions, by age group.}
\begin{center}
\begin{tabular}{|c|c|c|c|c|c|c|}
\hline
\multicolumn{2}{|c|}{} & 18-25 & 26-35 & 36-45 & 46-55 & 56-65 \\ \hline
 {1} & Yes & 94,93\% & 94,38\% & 98,08\% & 86,21\% & 81,25\% \\ \cline{2-7} 
 & No & 5,07\% & 5,62\% & 1,92\% & 13,79\% & 18,75\% \\ \hline
 {2} & Yes & 90,78\% & 94,38\% & 84,62\% & 75,86\% & 81,25\% \\ \cline{2-7} 
 & No & 9,22\% & 5,62\% & 15,38\% & 24,14\% & 18,75\% \\ \hline
 {3} & Yes & 71,43\% & 72,47\% & 71,15\% & 58,62\% & 75,00\% \\ \cline{2-7} 
 & No & 28,57\% & 27,53\% & 28,85\% & 41,38\% & 25,00\% \\ \hline
 {4} & Yes & 87,10\% & 85,39\% & 82,69\% & 65,52\% & 68,75\% \\ \cline{2-7} 
 & No & 12,90\% & 14,61\% & 17,31\% & 34,48\% & 31,25\% \\ \hline
 {5} & Yes & 90,78\% & 85,96\% & 84,62\% & 86,21\% & 87,50\% \\ \cline{2-7} 
 & No & 9,22\% & 14,04\% & 15,38\% & 13,79\% & 12,50\% \\ \hline
 {6} & Yes & 90,78\% & 87,08\% & 82,69\% & 86,21\% & 75,00\% \\ \cline{2-7} 
 & No & 9,22\% & 12,92\% & 17,31\% & 13,79\% & 25,00\% \\ \hline
 {7} & Yes & 16,13\% & 13,48\% & 11,54\% & 6,90\% & 0,00\% \\ \cline{2-7} 
 & No & 83,87\% & 86,52\% & 88,46\% & 93,10\% & 100,00\% \\ \hline
\end{tabular}
\label{tabp17eidade}
\end{center}
\end{table}


On table \ref{tabpsexo} the answers for questions 1 to 7 are segmented by sex. Once again, the results seem uniformly distributed, with the differences falling within the error margin of the survey.

\begin{table}[htb]
\caption{Answer of the objective questions, by sex.}
\begin{center}
\begin{tabular}{|c|c|c|c|}
\hline
\multicolumn{2}{|c|}{} & Male & Female \\ \hline
 {1} & Yes & 93,08\% & 95,17\% \\ \cline{2-4} 
 & No & 6,92\% & 4,83\% \\ \hline
 {2} & Yes & 89,97\% & 90,82\% \\ \cline{2-4} 
 & No & 10,03\% & 9,18\% \\ \hline
 {3} & Yes & 69,55\% & 73,91\% \\ \cline{2-4} 
 & No & 30,45\% & 26,09\% \\ \hline
 {4} & Yes & 82,35\% & 86,96\% \\ \cline{2-4} 
 & No & 17,65\% & 13,04\% \\ \hline
 {5} & Yes & 88,24\% & 87,92\% \\ \cline{2-4} 
 & No & 11,76\% & 12,08\% \\ \hline
 {6} & Yes & 85,81\% & 90,82\% \\ \cline{2-4} 
 & No & 14,19\% & 9,18\% \\ \hline
 {7} & Yes & 13,84\% & 13,04\% \\ \cline{2-4} 
 & No & 86,16\% & 86,96\% \\ \hline
\end{tabular}
\label{tabpsexo}
\end{center}
\end{table}


On table \ref{tab7renda} the answers for questions 1 to 7 are segmented by income bracket. While for questions 5 and 6 the results seem to not be affected by the familiar income of the interviewee, question 1 to 4 present a slight indication, considering the error margin of the survey, that people with less income are more prone to accept the intervention of a EMS in order to obtain a reduction on their energy bill. Conversely, by the results of 7, people with higher income are more inclined to change their mind about load shedding when receiving financial compensation for that.

\begin{table}[htb]
\caption{Answer of the objective questions, by income bracket.}
\begin{center}
\begin{tabular}{|c|c|c|c|c|c|c|}
\hline
\multicolumn{2}{|c|}{} & E & D & C & B & A \\ \hline
 {1} & Yes & 96,88\% & 94,34\% & 93,66\% & 94,25\% & 86,84\% \\ \cline{2-7} 
 & No & 3,13\% & 5,66\% & 6,34\% & 5,75\% & 13,16\% \\ \hline
{2} & Yes & 95,31\% & 88,68\% & 91,55\% & 89,66\% & 84,21\% \\ \cline{2-7} 
 & No & 4,69\% & 11,32\% & 8,45\% & 10,34\% & 15,79\% \\ \hline
 {3} & Yes & 78,13\% & 83,02\% & 66,20\% & 65,52\% & 42,11\% \\ \cline{2-7} 
 & No & 21,88\% & 16,98\% & 33,80\% & 34,48\% & 57,89\% \\ \hline
 {4} & Yes & 87,50\% & 89,94\% & 86,62\% & 75,86\% & 65,79\% \\ \cline{2-7} 
 & No & 12,50\% & 10,06\% & 13,38\% & 24,14\% & 34,21\% \\ \hline
 {5} & Yes & 87,50\% & 87,42\% & 90,14\% & 88,51\% & 84,21\% \\ \cline{2-7} 
 & No & 12,50\% & 12,58\% & 9,86\% & 11,49\% & 15,79\% \\ \hline
 {6} & Yes & 93,75\% & 86,79\% & 88,73\% & 82,76\% & 89,47\% \\ \cline{2-7} 
 & No & 6,25\% & 13,21\% & 11,27\% & 17,24\% & 10,53\% \\ \hline
 {7} & Yes & 9,38\% & 7,55\% & 21,13\% & 11,49\% & 23,68\% \\ \cline{2-7} 
 & No & 90,63\% & 92,45\% & 78,87\% & 88,51\% & 76,32\% \\ \hline
\end{tabular}
\label{tab7renda}
\end{center}
\end{table}

The overall results indicates a positive view of the intervention of a EMS on the way people consume energy if that means stepping towards a more efficient and cost effective consumption.


\subsection{Clustering}

Aiming to enable the usage of local data for QoE-aware DM action, for each of the equipment that were inquired about in the  survey, clustering was performed using the k-means algorithm with k being chosen accordingly to the elbow method result. This allowed for representative curves to be leveraged. Such curves describe the average behaviors and could, thus, be used as reference values for QoE-aware DM policies.

The equipment that allow for both anticipation and delay had their responses split in two, so the anticipation and delay problem were considered separately for fidelity of clustering purposes
. That way a complete curve for a specific user is supposed to be formed by the combination of an anticipation and a delay curve.


As the elbow method results were similar in all equipment they are shown just for the first equipment. Considering the obtained results, a k equals 5 was deemed appropriate for all of the appliances.

\begin{figure}[htbp]
\centerline{\includegraphics[scale=.6]{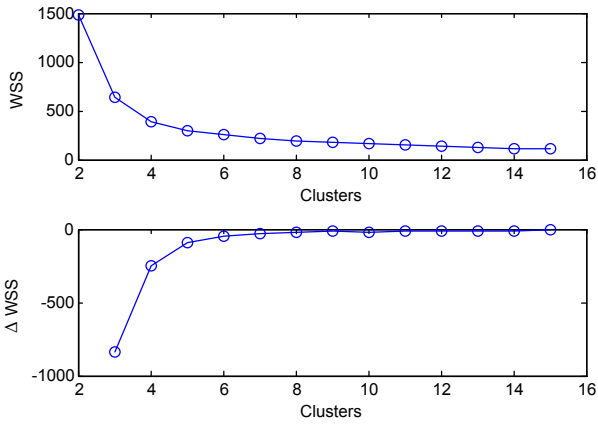}}
\caption{Elbow method results for the washer machine anticipation.}
\label{QoE01}
\end{figure}

The figure \ref{QoE01} presents the result of the elbow method for the anticipation curve of the washer machine: the within-cluster sum of square (WSS) and the increment of WSS due to increasing k.

It is also important to highlight that the users were asked to rate their discomfort level with the described actions, so, the values presented below represent the Complementary QoE (CQoE), as a low value represent a satisfactory experience and a high value a unsatisfactory experience.

The results of the obtained centroids of each cluster are shown on table \ref{tabantic} for the anticipation of each equipment. The sequential number of each cluster is set on the first column and the CQoE score for each amount of time are given on the subsequent columns, with 1 representing low discomfort and 5 representing intense discomfort.

\begin{table}[htb]
\caption{Resulting clusters for the anticipation of the starting time of washing machine, clothes dryer and dishwasher.}
\begin{center}
\begin{tabular}{|c|c|c|c|c|c|c|}
\hline
        & -4h    & -3h    & -2h    & -1h    & -0.5h  & +0h    \\ \hline
        \multicolumn{7}{|c|}{washing machine} \\ \hline
1       & 1.0483 & 1.0000 & 1.0000 & 1.0000 & 1.0000 & 1.0000 \\ \hline
2       & 4.7656 & 4.3281 & 3.6094 & 2.0781 & 1.2813 & 1.0000 \\ \hline
3       & 2.5000 & 2.0294 & 1.3382 & 1.1176 & 1.0735 & 1.0000 \\ \hline
4       & 3.6622 & 3.1216 & 2.1622 & 1.3514 & 1.1757 & 1.0000 \\ \hline
5       & 4.9143 & 4.9143 & 4.9143 & 4.9143 & 4.9143 & 1.0000 \\ \hline
        \multicolumn{7}{|c|}{clothes dryer} \\ \hline
1       & 2.9670 & 2.4835 & 1.7473 & 1.1538 & 1.0000 & 1.0000 \\ \hline
2       & 1.0140 & 1.0000 & 1.0000 & 1.0000 & 1.0000 & 1.0000 \\ \hline
3       & 4.7344 & 4.1563 & 3.2188 & 1.9219 & 1.2031 & 1.0000 \\ \hline
4       & 2.5238 & 2.4762 & 2.4286 & 2.4286 & 2.4286 & 1.0000 \\ \hline
5       & 5.0000 & 5.0000 & 4.9828 & 4.8621 & 4.7931 & 1.0000 \\ \hline
        \multicolumn{7}{|c|}{dishwasher} \\ \hline
1       & 1.0238 & 1.0000 & 1.0000 & 1.0000 & 1.0000 & 1.0000 \\ \hline
2       & 2.4902 & 2.0588 & 1.3529 & 1.1961 & 1.1373 & 1.0000 \\ \hline
3       & 4.9589 & 4.9589 & 4.9452 & 4.9452 & 4.8630 & 1.0000 \\ \hline
4       & 3.6607 & 3.1964 & 2.2321 & 1.4821 & 1.2321 & 1.0000 \\ \hline
5       & 4.7759 & 4.3276 & 3.5517 & 2.0862 & 1.3103 & 1.0000 \\ \hline

\end{tabular}
\label{tabantic}
\end{center}
\end{table}

The results of the centroids for the postponing of each equipment are shown on table \ref{tabpost}.

\begin{table}[htb]
\caption{Resulting clusters for the postponing of the starting time of washing machine, clothes dryer and dishwasher.}
\begin{center}
\begin{tabular}{|c|c|c|c|c|c|c|}
\hline
        & +0h    & +0.5h  & +1h    & +2h    & +3h    & +4h    \\ \hline
        \multicolumn{7}{|c|}{washing machine} \\ \hline
1       & 1.0000 & 1.1939 & 1.3571 & 2.0918 & 3.1735 & 3.7959 \\ \hline
2       & 1.0000 & 1.0000 & 1.0000 & 1.0000 & 1.0000 & 1.0497 \\ \hline
3       & 1.0000 & 1.0923 & 1.1231 & 1.3231 & 2.0462 & 2.5385 \\ \hline
4       & 1.0000 & 4.8710 & 4.9032 & 4.9355 & 4.9355 & 4.9355 \\ \hline
5       & 1.0000 & 1.3151 & 2.1507 & 3.4247 & 4.3836 & 4.8767 \\ \hline
        \multicolumn{7}{|c|}{clothes dryer} \\ \hline
1       & 1.0000 & 1.1385 & 1.8000 & 2.8000 & 3.9846 & 4.8000 \\ \hline
2       & 1.0000 & 1.0000 & 1.0000 & 1.0000 & 1.0505 & 1.0657 \\ \hline
3       & 1.0000 & 4.9800 & 5.0000 & 5.0000 & 5.0000 & 5.0000 \\ \hline
4       & 1.0000 & 1.2642 & 1.3774 & 1.9623 & 2.6132 & 3.0472 \\ \hline
5       & 1.0000 & 1.9310 & 2.7931 & 4.4828 & 4.7586 & 4.8621 \\ \hline
        \multicolumn{7}{|c|}{dishwasher} \\ \hline
1       & 1.0000 & 1.1250 & 1.1607 & 1.3036 & 2.0357 & 2.6429 \\ \hline
2       & 1.0000 & 1.3768 & 2.2319 & 3.5652 & 4.3478 & 4.8261 \\ \hline
3       & 1.0000 & 1.0000 & 1.0000 & 1.0000 & 1.0000 & 1.0348 \\ \hline
4       & 1.0000 & 1.2097 & 1.4355 & 2.2258 & 3.2419 & 3.7742 \\ \hline
5       & 1.0000 & 4.8833 & 4.9667 & 4.9667 & 4.9667 & 4.9667 \\ \hline

\end{tabular}
\label{tabpost}
\end{center}
\end{table}

The centroids for the curtailable equipaments are presented on table \ref{tabcurt}. Appendix A contains the curves for the centroids leveraged for each equipment.

\begin{table}[htb]
\caption{Resulting clusters for the intervention on the temperature of heating, air conditioning and shower, and extension on the electric vehicle loading time.}
\begin{center}
\begin{tabular}{|c|c|c|c|c|c|c|}
\hline

        \multicolumn{7}{|c|}{heating} \\ \hline
        &-0$^\circ$C&-1$^\circ$C&-2$^\circ$C&-3$^\circ$C&-4$^\circ$C&-5$^\circ$C\\ \hline
1       & 1.0000    & 1.1282    & 1.8462    & 3.1795    & 4.2436    & 4.9231 \\ \hline
2       & 1.0000    & 1.0000    & 1.0000    & 1.0000    & 1.0000    & 1.0926 \\ \hline
3       & 1.0000    & 1.0423    & 1.0845    & 1.4225    & 2.0000    & 2.5352 \\ \hline
4       & 1.0000    & 1.3231    & 1.4000    & 2.1538    & 3.0462    & 3.7538 \\ \hline
5       & 1.0000    & 4.9365    & 4.9762    & 4.9841    & 5.0000    & 5.0000 \\ \hline
        \multicolumn{7}{|c|}{air conditioning} \\ \hline
        &+0$^\circ$C&+1$^\circ$C&+2$^\circ$C&+3$^\circ$C&+4$^\circ$C&+5$^\circ$C\\ \hline
1       & 1.0000    & 1.0233    & 1.7093    & 3.1395    & 4.1628    & 4.9070 \\ \hline
2       & 1.0000    & 4.9840    & 4.9947    & 4.9947    & 4.9947    & 4.9947 \\ \hline
3       & 1.0000    & 1.0351    & 1.0351    & 1.1579    & 1.4912    & 1.5789 \\ \hline
4       & 1.0000    & 1.0300    & 1.0900    & 1.6800    & 2.6800    & 3.4700 \\ \hline
5       & 1.0000    & 2.0556    & 3.2222    & 4.1111    & 4.5556    & 4.8889 \\ \hline
        \multicolumn{7}{|c|}{shower} \\ \hline
        &-0$^\circ$C&-1$^\circ$C&-2$^\circ$C&-3$^\circ$C&-4$^\circ$C&-5$^\circ$C\\ \hline
1       & 1.0000    & 1.1833    & 1.3167    & 2.1167    & 3.0500    & 3.7000 \\ \hline
2       & 1.0000    & 4.9918    & 4.9918    & 4.9959    & 4.9959    & 4.9959 \\ \hline
3       & 1.0000    & 1.1791    & 1.9552    & 3.1493    & 4.1045    & 4.8955 \\ \hline
4       & 1.0000    & 1.0351    & 1.0351    & 1.3158    & 2.0877    & 2.5965 \\ \hline
5       & 1.0000    & 1.0000    & 1.0000    & 1.0000    & 1.0000    & 1.2105 \\ \hline
        \multicolumn{7}{|c|}{electric vehicle} \\ \hline
        & +0h    & +0.5h  & +1h    & +2h    & +3h    & +4h    \\ \hline
1       & 1.0000 & 4.9851 & 4.9963 & 4.9963 & 4.9963 & 4.9963 \\ \hline
2       & 1.0000 & 1.1111 & 1.6444 & 2.8889 & 4.2000 & 4.8889 \\ \hline
3       & 1.0000 & 1.1667 & 1.3750 & 2.1667 & 3.0208 & 3.6042 \\ \hline
4       & 1.0000 & 1.0526 & 1.0789 & 1.3158 & 2.0789 & 2.8158 \\ \hline
5       & 1.0000 & 1.0000 & 1.0000 & 1.0000 & 1.0000 & 1.0833 \\ \hline

\end{tabular}
\label{tabcurt}
\end{center}
\end{table}

The number of responses belonging to each cluster are shown on table \ref{tabmembers}.

\begin{table}[htb]
\caption{Number of members belonging to each cluster.}
\begin{center}
\begin{tabular}{|c|c|c|c|c|c|}
\hline
Cluster &  1  &  2  &  3  &  4  &  5   \\ \hline
 \multicolumn{6}{|c|}{washing machine anticipation} \\ \hline
Members & 207 & 64  & 68  & 74  & 35 \\ \hline
 \multicolumn{6}{|c|}{clothes dryer anticipation} \\ \hline
 Members & 91  & 214 & 64  & 21  & 58 \\ \hline
 \multicolumn{6}{|c|}{dishwasher anticipation} \\ \hline
 Members & 210 & 51  & 73  & 59  & 58 \\ \hline
  \multicolumn{6}{|c|}{washing machine postponing} \\ \hline
   Members & 65  & 181 & 98  & 31  & 73 \\ \hline
 \multicolumn{6}{|c|}{clothes dryer postponing} \\ \hline
Members & 65  & 198 & 50  & 106 & 29 \\ \hline
 \multicolumn{6}{|c|}{dishwasher postponing} \\ \hline
 Members & 56  & 69  & 201 & 62  & 60 \\ \hline
 \multicolumn{6}{|c|}{heating curtailment} \\ \hline
 Members & 78  & 108 & 71  & 65  & 126 \\ \hline
 \multicolumn{6}{|c|}{air conditioning curtailment} \\ \hline
 Members & 86  & 187 & 57  & 100 & 18 \\ \hline
 \multicolumn{6}{|c|}{shower curtailment} \\ \hline
 Members & 60  & 254 & 67  & 57  & 19 \\ \hline
 \multicolumn{6}{|c|}{electric vehicle curtailment} \\ \hline
 Members & 269 & 45  & 48  & 38  & 48 \\ \hline
  
\end{tabular}
\label{tabmembers}
\end{center}
\end{table}

Concerning the washing machine, the most populated clusters are the ones representing high willingness of anticipating or postponing the usage of the appliance despite the length of the time shift. An equivalent amount of people, 206 for anticipation (summing clusters 2, 3 and 4) and 236 for postponing (clusters 1, 3 and 5), would allow some shift but with increasing discontent, with varying rate of discomfort growth. For these users a QoE-aware EMS would be most suited to deploy demand management actions. On the other hand, less than 8 \% of the respondents would not welcome any time shift on the usage of their washing machine. 

Similar results are found for the the clothes dryer and dishwasher, but, for these appliances, at most 13 \% of the users would not accept interference on the starting time of the appliances. Nevertheless, the remaining homes would be suitable for installation of a QoE-aware EMS.

The results for postponing the washing machine, the clothes dryer and the dishwasher were expected, as changes in the starting time of these appliances should not interfere heavily with the routine of the users. Anticipating could be deemed as more bothersome, but the results do not show much difference in the numbers of people unwilling of changing the usage time.

For the curtailing of the power of the heating, the second cluster, roughly one fourth of the answers, represents users that would agree to have their ambient temperature decreased by up to five degrees if that causes a reduction in their energy tariff. The fifth group, with more than 28 \% of the users, however, gathers those who would not agree with changes on the desired temperature for heating. The remaining three clusters represent users that would allow some reduction on temperature but that would imply in increasing discontent, with varying rate of discomfort growth rates.

For the other curtailable loads the numbers of users not willing to tolerate any curtailment grows to 41.74 \%, 56.70 \% and 60.04 \% concerning the air conditioning, shower and electric vehicle charge. The higher numbers were expected as the objective questions had already pointed out that the interviewees were less receptive with curtailing loads than with simply shifting their starting time. That said, the remaining market for QoE-aware EMS able to curtail the power consumption of these appliances is still relevant.

\subsection{General Shiftable Load}

While individual curves for equipments are useful for a EMS acting on each house, a single curve that depicts the willingness of the consumer to shift load in time is useful to a microgrid central controller or even to a distribution network operator. In spite of that, an average curve of the shiftable loads was leveraged for each user. The clustering of such curves is presented below.



\begin{figure}[htbp]
\centerline{\includegraphics[scale=.6]{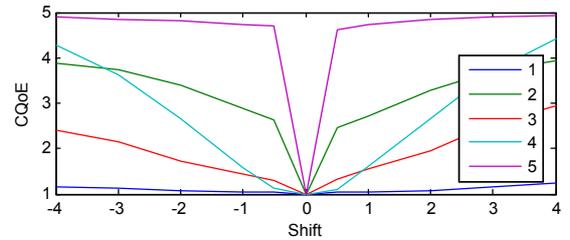}}
\caption{Curves for the average shiftable load.}
\label{QoE36}
\end{figure}

The result of the obtained centroids of each cluster are shown on figure \ref{QoE36}.

The values for each curve are presented on table \ref{tabshift}. In this table the groups are presented on the columns.

\begin{table}[htb]
\caption{Resulting clusters for general shiftable loads.}
\begin{center}
\begin{tabular}{|c|c|c|c|c|c|}
\hline
      &     1     &    2      &    3      &   4       &   5 \\  \hline   
-4h    & 1.1657    & 3.8918    & 2.3922    & 4.2843    &4.8939 \\ \hline
-3h    & 1.1229    & 3.7403    & 2.1405    & 3.6176    &4.8485\\ \hline
-2h    & 1.0726    & 3.4069    & 1.7288    & 2.6569    &4.8030\\ \hline
-1h    & 1.0372    & 2.8788    & 1.4477    & 1.5735    &4.7424\\ \hline
-0.5h  & 1.0335    & 2.6277    & 1.3039    & 1.1176    &4.7121\\ \hline
+0h    & 1.0000    & 1.0000    & 1.0000    & 1.0000    &1.0000\\ \hline
+0.5h  & 1.0317    & 2.4589    & 1.3170    & 1.0980    &4.6212\\ \hline
+1h    & 1.0372    & 2.7100    & 1.5425    & 1.6029    &4.7424\\ \hline
+2h    & 1.0819    & 3.2814    & 1.9608    & 2.6520    &4.8333\\ \hline
+3h    & 1.1639    & 3.7186    & 2.5490    & 3.7108    &4.8939\\ \hline
+4h    & 1.2272    & 3.9437    & 2.9379    & 4.4118    &4.9242\\ \hline
\end{tabular}
\label{tabshift}
\end{center}
\end{table}

The number of responses belonging to each cluster are presented on table \ref{tabsh}.

\begin{table}[htb]
\caption{Number of members on each cluster for the electric car battery charging.}
\begin{center}
\begin{tabular}{|c|c|c|c|c|c|}
\hline
Cluster &  1  &  2  &  3  &  4  &  5   \\ \hline
Members & 179 & 77  & 102 & 68  & 22 \\ \hline
\end{tabular}
\label{tabsh}
\end{center}
\end{table}



The first cluster, which is the most numerous group, represents users who would agree with anticipations or delays on their shiftable loads with minimal discomfort, if that shift meant a reduction on their energy bill. The fifth cluster, conversely, represents the users that would not allow intervention on their shiftable loads, for such intervention would mean severe discomfort. The other three clusters portray users that would allow some interference, whose perceived annoyance would increase with different rates as the shift period increase. The result agrees with what was expected from the objective questions.

\subsection{General Curtailable Load}



\begin{figure}[htbp]
\centerline{\includegraphics[scale=.6]{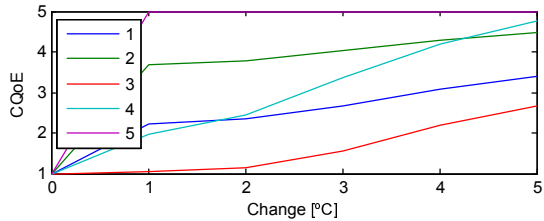}}
\caption{Curves for the average curtailable load.}
\label{QoE39}
\end{figure}

Following similar motivations, an average curve for the curtailable loads related to thermal loads was leveraged for each user. The clustering of these curves is presented on figure \ref{QoE39}.

The values for each curve are presented on table \ref{tabshed}.

\begin{table}[htb]
\caption{Resulting clusters for general curtailable thermal load, considering a change in the temperature of a given value.}
\begin{center}
\begin{tabular}{|c|c|c|c|c|c|c|}
\hline
        &0$^\circ$C&1$^\circ$C&2$^\circ$C&3$^\circ$C&4$^\circ$C&5$^\circ$C\\ \hline
1       & 1.0000 & 2.2075 & 2.3396 & 2.6824 & 3.0881 & 3.4057 \\ \hline
2       & 1.0000 & 3.6667 & 3.7885 & 4.0466 & 4.2796 & 4.4624 \\ \hline
3       & 1.0000 & 1.0449 & 1.1273 & 1.5618 & 2.1798 & 2.6779 \\ \hline
4       & 1.0000 & 1.9759 & 2.4498 & 3.3735 & 4.1888 & 4.7550 \\ \hline
5       & 1.0000 & 4.9870 & 4.9913 & 4.9957 & 5.0000 & 5.0000 \\ \hline
\end{tabular}
\label{tabshed}
\end{center}
\end{table}

The number of responses belonging to each cluster are presented on table \ref{tabsh}.

\begin{table}[htb]
\caption{Number of members on each cluster for the general curtailable thermal load.}
\begin{center}
\begin{tabular}{|c|c|c|c|c|c|}
\hline
Cluster &  1  &  2  &  3  &  4  &  5   \\ \hline
Members & 106 & 93  & 89  & 83  & 77 \\ \hline
\end{tabular}
\label{tabsh}
\end{center}
\end{table}



For this particular set of clusters, all of them have quite similar number of members. The fifth cluster, which is the least numerous group, encompasses users that would not feel comfortable with intervention on their curtailable load, perceiving high discomfort if that would occur. The third group is the most prone to accept DM actions on their thermal loads, but even then would perceive some discomfort degree depending on how severe was the intervention. The other three clusters portray users that would allow some interference, whose perceived annoyance would increase with different rates as the shift period increase.

\section{Conclusions and Policy Implications}

The present work has discussed the results of a survey concerning QoE-aware demand management. It was conducted at Federal University of Santa Catarina, Forian\'{o}polis/Brazil, and surveyed students, faculty members and local community members. The online questionnaire answered by the interviewees was composed of two main parts: an identification section and a technical section. While the identification section was employed to characterize the population socio-economical aspects, the technical section aimed to obtain the interviewees' perceptions about interacting with energy management systems on their houses, and rate the impact of DM actions on a QoE frame.

The interviewees were mostly favorable to the perspective of having a smart system interfering on their energy consumption, if that meant a reduction on their energy bill. This disposition aligns with the crescent effort that has been put on research of EMSs for smart houses. In particular, for the Brazilian case, the legislation is leaning towards a  scenario where it will be more profitable to allign the load curve with the generation curve. EMSs capable of performing changes on the consumption profile will be very welcome and thus stand as a relevant study topic. On this matter, QoE-aware DM schemes seems to be a promising alternative, as it allows the EMS to consider both financial and satisfaction issues when adopting DM actions. On that note, this work delivered a set of QoE curves that represent the annoyance perceived by the users when subjected to certain DM actions related to domestic appliances. For each equipment the responses were clustered in order to portray standard behaviors that can be used as reference for QoE-aware DM on EMS developed to address the local market.

Additionally, the significant number of positive answers to the question on the EMS acting on the artificial lighting shows that there are an opening for research on systems that account for the natural lighting when dimming the artificial light sources, aiming to keep the illumination at adequate levels. The same conclusion can be drawn on the high number of favorable answers to the question on the bath duration.

Furthermore, considering the present opening in the regulation for the pilot demand response program and the answers concerning remunerated load shedding, a gap for more studies on the feasibility and impacts of a future evolution of the regulation to allow Group B consumers to band together in order to participate on demand response programs as virtual power plants. On that note, QoE metrics could also be integrated in such approach.

\begin{ack}
The authors thank CNPq for projects 401126/2014-5, 303702/2011-7 and 305785/2015-0.
\end{ack}

\bibliography{ifacconf}             
\appendix
\section{Graphical Presentation of the Obtained Clusters}


\begin{figure}[htbp]
\centerline{\includegraphics[scale=.6]{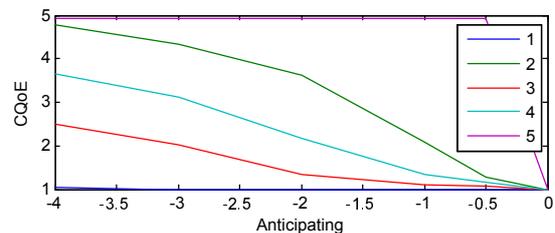}}
\caption{Curves for the washing machine anticipation.}
\label{QoE02}
\end{figure}


\begin{figure}[htbp]
\centerline{\includegraphics[scale=.6]{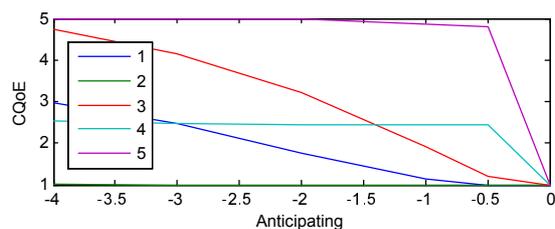}}
\caption{Curves for the clothes dryer anticipation.}
\label{QoE08}
\end{figure}


\begin{figure}[htbp]
\centerline{\includegraphics[scale=.6]{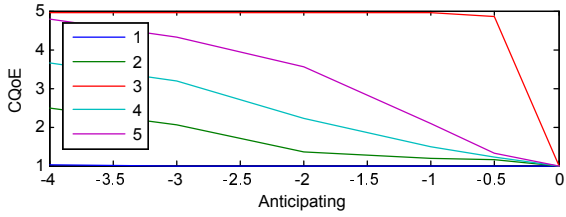}}
\caption{Curves for the dishwasher anticipation.}
\label{QoE14}
\end{figure}


\begin{figure}[htbp]
\centerline{\includegraphics[scale=.6]{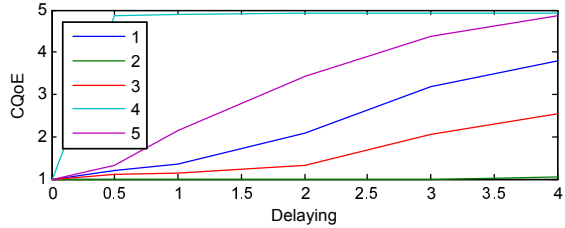}}
\caption{Curves for the washing machine postponing.}
\label{QoE05}
\end{figure}


\begin{figure}[htbp]
\centerline{\includegraphics[scale=.6]{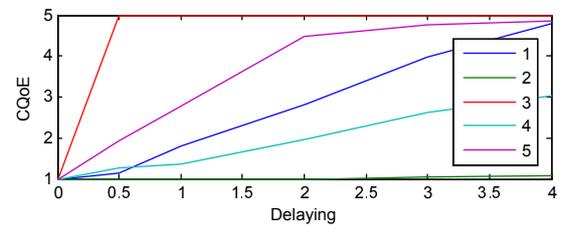}}
\caption{Curves for the clothes dryer delay.}
\label{QoE11}
\end{figure}


\begin{figure}[htbp]
\centerline{\includegraphics[scale=.6]{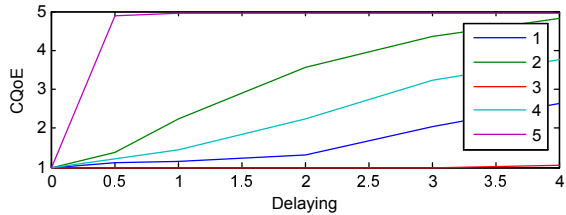}}
\caption{Curves  for the dishwasher postponing.}
\label{QoE17}
\end{figure}


\begin{figure}[htbp]
\centerline{\includegraphics[scale=.6]{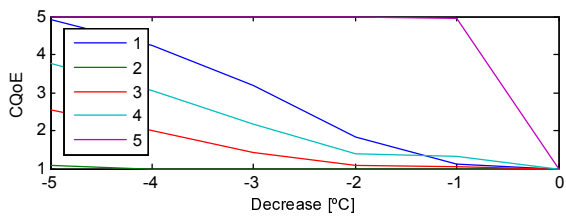}}
\caption{Curves for the heating curtailing.}
\label{QoE20}
\end{figure}


\begin{figure}[htbp]
\centerline{\includegraphics[scale=.6]{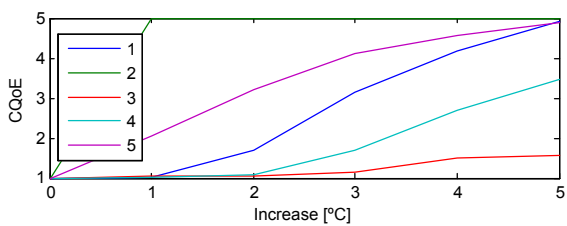}}
\caption{Curves for the air conditioning curtailing.}
\label{QoE23}
\end{figure}


\begin{figure}[htbp]
\centerline{\includegraphics[scale=.6]{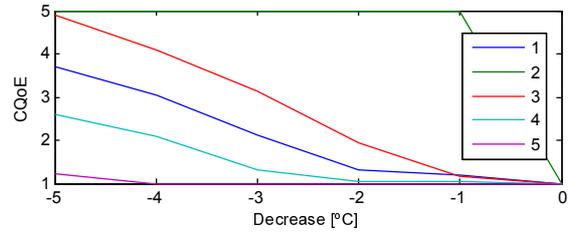}}
\caption{Curves for the electric shower curtailing.}
\label{QoE26}
\end{figure}


\begin{figure}[htbp]
\centerline{\includegraphics[scale=.6]{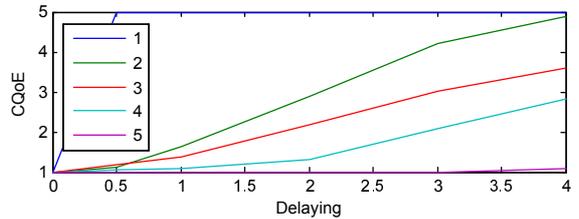}}
\caption{Curves for the electric car charging delaying.}
\label{QoE29}
\end{figure}


\end{document}